\pdfoutput=1
\documentclass[sigconf]{acmart}
\renewcommand\footnotetextcopyrightpermission[1]{} 
\usepackage{multirow}
\usepackage{mathrsfs}
\usepackage{amsfonts}
\usepackage{mathtools}
\usepackage[normalem]{ulem}
\useunder{\uline}{\ul}{}
\graphicspath{ {./images/} }

\AtBeginDocument{%
  \providecommand\BibTeX{{%
    \normalfont B\kern-0.5em{\scshape i\kern-0.25em b}\kern-0.8em\TeX}}}

\copyrightyear{2020}
\acmYear{2020}
\setcopyright{acmcopyright}\acmConference[SIGIR '20]{Proceedings of the 43rd International ACM SIGIR Conference on Research and Development in Information Retrieval}{July 25--30, 2020}{Virtual Event, China}
\acmBooktitle{Proceedings of the 43rd International ACM SIGIR Conference on Research and Development in Information Retrieval (SIGIR '20), July 25--30, 2020, Virtual Event, China}
\acmPrice{15.00}
\acmDOI{10.1145/3397271.3401240}
\acmISBN{978-1-4503-8016-4/20/07}

\begin{document}
\fancyhead{}

\title{MRIF: Multi-resolution Interest Fusion for Recommendation}

\author{Shihao Li}
\affiliation{%
	\institution{Alibaba Inc}
	\streetaddress{Wenyi Street 969}
	\city{Hangzhou}
	\country{China}
	\postcode{311121}
}
\email{shihao.lsh@alibaba-inc.com}

\author{Dekun Yang}
\affiliation{%
	\institution{Alibaba Inc}
	\streetaddress{Wenyi Street 969}
	\city{Hangzhou}
	\country{China}
	\postcode{311121}
}
\email{dekun.ydk@alibaba-inc.com}

\author{Bufeng Zhang}
\affiliation{%
	\institution{Alibaba Inc}
	\streetaddress{Wenyi Street 969}
	\city{Hangzhou}
	\country{China}
	\postcode{311121}
}
\email{feitong@alibaba-inc.com}


\renewcommand{\shortauthors}{Li, et al.}

\begin{abstract}
  The main task of personalized recommendation is capturing users' interests based on their historical behaviors. Most of recent advances in recommender systems mainly focus on modeling users' preferences accurately using deep learning based approaches. There are two important properties of users' interests, one is that users' interests are dynamic and evolve over time, the other is that users' interests have different resolutions, or temporal-ranges to be precise, such as long-term and short-term preferences. Existing approaches either use Recurrent Neural Networks (RNNs) to address the drifts in users' interests without considering different temporal-ranges, or design two different networks to model long-term and short-term preferences separately. This paper presents a multi-resolution interest fusion model (MRIF) that takes both properties of users' interests into consideration. The proposed model is capable to capture the dynamic changes in users' interests at different temporal-ranges, and provides an effective way to combine a group of multi-resolution user interests to make predictions. Experiments show that our method outperforms state-of-the-art recommendation methods consistently.
\end{abstract}



\maketitle

\section{Introduction}
In recent years, recommender systems have been evolving fast. As deep learning methods achieve state-of-the-art performances in a lot of fields such as computer vision and natural language processing, several deep learning based recommendation methods are developed by extending the traditional collaborative filtering techniques \cite{he2017neural}. 

However, users' interests are dynamic and change over time, which are hard to express by simple factorization approaches. The sequential recommender has attracted much attention recently due to its ability to capture users' intents based on the order and relation of user behaviors. GRU4Rec \cite{hidasi2015gru4rec} uses GRU-based RNN to extract information from user interaction sequences. CASER \cite{tang2018personalized} embeds an item sequence into an image and learns sequential patterns via horizontal and vertical convolutional filters.

Although these sequential recommenders manage to extract main user interests through sequential user interactions, the evolution process and resolution of user interests are lost. There are two important properties of users' interests, one is that users' interests are dynamic and evolve over time, the other is that users' interests have different resolutions, such as long-term and short-term preferences. In this paper, we introduce multi-resolution interest fusion model (MRIF) composed of interest extraction layer, interest aggregation layer, and attentional fusion structure, which addresses the problem of extracting users' preferences at different temporal-ranges and combining multi-resolution interests effectively. The main contributions are:
\begin{itemize}
	\item We design a new network structure to model the dynamic changes and different temporal-ranges of users' interests, which yields more accurate prediction results than extracting main interest directly from interaction sequences.
	\item We propose three different aggregators, namely mean aggregator, max aggregator, and attentional aggregator, to capture users' interests at different temporal-ranges.
	\item We conduct experiments on two different datasets. The experiment results show that our method outperforms other state-of-the-art methods consistently.
\end{itemize}
\begin{figure*}[t!]
	\includegraphics[width=0.95\textwidth]{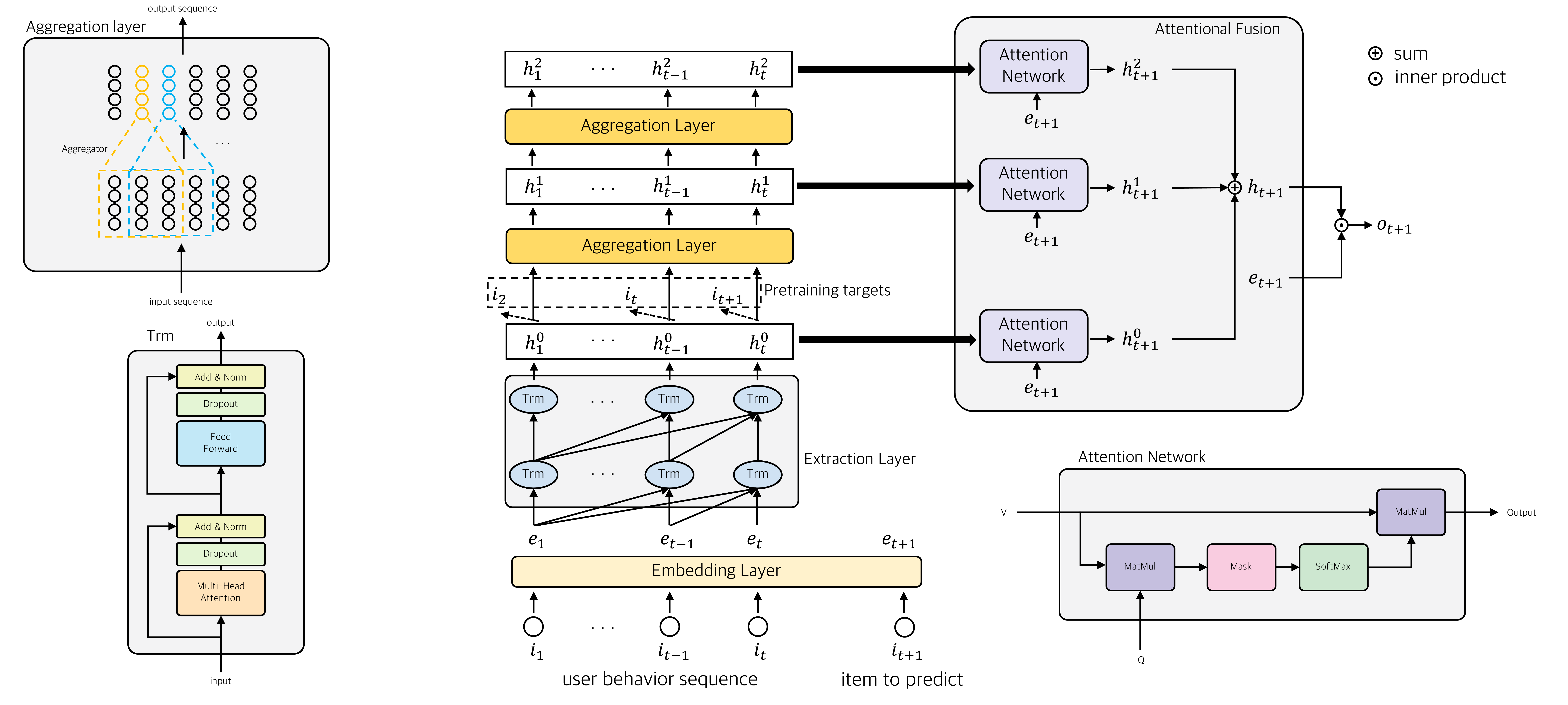}
	\caption{The structure of MRIF. User behaviors go through embedding layer and interest extraction layer to obtain instantaneous user interests, which are then fed into stacking aggregation layers to extract short term and long term preferences. Instantaneous, short-term, and long-term user interests are then combined through attention network to make predictions.
	}
	\label{fig:mrif}
\end{figure*}
\section{Proposed Method}
In this section, we introduce our MRIF model in detail. As is shown in Fig.~\ref{fig:mrif}, the proposed model is composed of three main parts, which are interest extraction layer, interest aggregation layer, and attentional fusion structure. Interest extraction layer extracts instantaneous user interests from embedded behavior sequences. Interest aggregation layer captures users' interests at different temporal-ranges. Attentional fusion structure combines users' interests using attentional mechanisms to make predictions.

\subsection{Interest Extraction Layer}
Interest extraction layer is positioned above embedding layer, which represents each item using a fixed length latent vector. For an user's item sequence $\vec{i}=(i_1,i_2,\cdots,i_n)$, the embedding layer project the sequence into an embedding matrix $\mathbf{E} \in \mathbb{R}^{n\times d}$. The embedding matrix is the sum of item embedding $\mathbf{M} \in \mathbb{R}^{n\times d}$ and positional embedding $\mathbf{P} \in \mathbb{R}^{n\times d}$. If the length of user's item sequence is less than $n$, zero item embeddings are appended.

The interest of a user at each step can be modeled as a hidden variable, which can not be observed directly but can be estimated by historical behaviors. Previous work uses Hidden Markov Model (HMM) to predict users' latent interests by maximizing probabilities of behavior sequences under hidden user interests \cite{sahoo2010hidden}. However, the states of HMM model are very limited and can not effectively express a vast space of user interests. 
DIEN \cite{zhou2019dien} chooses GRU-based RNN as user interest extractor, which is time-consuming for long sequences. Transformer network proposed in \cite{vaswani2017attention} relies on self-attention instead of recurrence, which is more efficient and achieves superior performance. We use transformer as our interest extractor and pre-train transformer network to predict the next item in sequence at each step. Transformer network is composed of two main parts, multi-head attention and feedforward network. Multi-head attention projects input sequence embedding $\mathbf{X}\in \mathbb{R}^{n \times d}$ into $h$ subspaces, then applies scaled dot-product attention function on each subspace:
\begin{equation}
\mathrm{MultiHead}(\mathbf{X}) = \mathrm{Concat}(\mathrm{head_1}(\mathbf{X}),\cdots,\mathrm{head_h}(\mathbf{X}))\mathbf{W}
\end{equation}
\begin{equation}
\mathrm{head_i}(\mathbf{X}) = \mathrm{Attention}(\mathbf{XW}_i^\mathbf{Q},\mathbf{XW}_i^K,\mathbf{XW}_i^V)
\end{equation}
\begin{equation}
\mathrm{Attention}(\mathbf{Q},\mathbf{K},\mathbf{V}) = \mathrm{softmax}(\frac{\mathbf{QK}^T}{\sqrt{d}})\mathbf{V}.
\end{equation}
Feedforward network applies two affine transforms and ReLU activation to adds nonlinearity:
\begin{equation}
\mathrm{FFN}(\mathbf{X}) = \mathrm{ReLU}(\mathbf{XW}_1+\mathbf{b}_1)\mathbf{W}_2+\mathbf{b}_2.
\end{equation}
The transformer network is built upon multi-head attention and feedforward network, where dropout, layer normalization, and residual connection is added.
The equation of transformer layer is as follows:
\begin{equation}
\mathrm{Trm}(\mathbf{X}) = \mathrm{LN}(\mathrm{Dropout}(\mathrm{FFN}(\mathrm{SA}(\mathbf{X})))+\mathrm{SA}(\mathbf{X}))
\end{equation}
\begin{equation}
\mathrm{SA}(\mathbf{X}) = \mathrm{LN}(\mathrm{Dropout}(\mathrm{MultiHead}(\mathbf{X}))+\mathbf{X})
\end{equation}
where $LN$ is layer normalization.
The instantaneous user interest $\mathbf{H}^0\in \mathbb{R}^{n \times d}$ is extracted by stacking two transformer layers:
\begin{equation}
\mathbf{H}^0 = \mathrm{Trm}(\mathrm{Trm}(\mathbf{E}))
\end{equation}
In order to capture instantaneous user interest at each step accurately, we pre-train the transformer network to predict the next behavior of a user at each step.

\begin{figure}[t!]
	\includegraphics[width=0.48\textwidth]{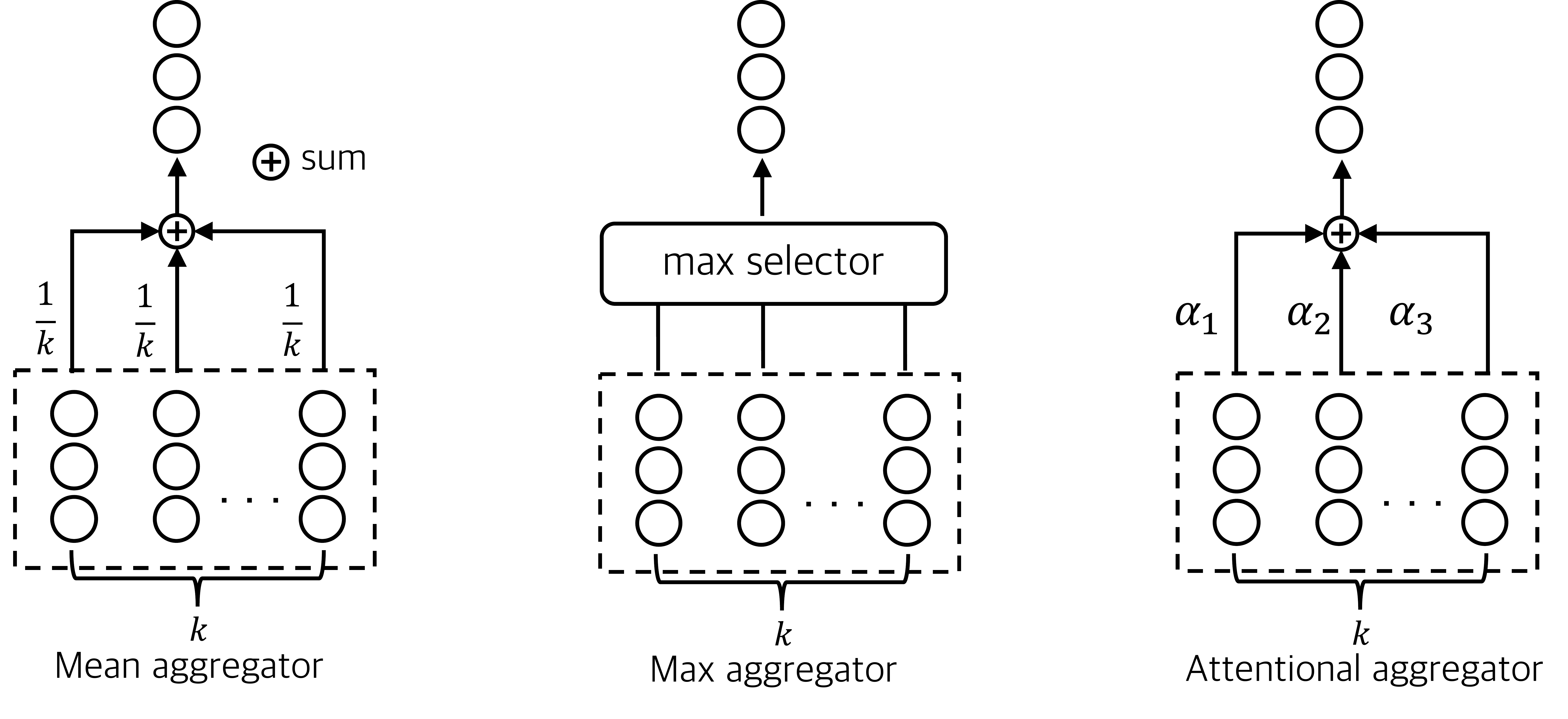}
	\caption{Three different types of aggregators. Mean aggregator takes the average of input embeddings. Max aggregator selects the input embedding with maximum norm from the embedding sequence. Attentional aggregator weights each input by an attentional score.
	}
\end{figure}

\subsection{Interest Aggregation Layer}
The purpose of interest aggregation layer is to inspect user interest at different temporal-ranges and form a group of multi-resolution user interests. Interest aggregation layer creates a sliding window with width $k=2w+1$ that moves along input embedding sequence one step at a time, and then applies aggregator to the windowed embeddings. Denote the output embedding sequence of aggregation layer $l$ as $\mathbf{H}^l\in \mathbf{R}^{n\times d}$, and embedding at step $j$ in $\mathbf{H}^l$ as $\mathbf{H}^l_j$, the output embedding of layer $l+1$ can computed as follows:
\begin{equation}
\mathbf{H}^{l+1}_i = \mathrm{Agg}([\mathbf{H}^{l}_{i-w},\mathbf{H}^{l}_{i-w+1},\cdots,\mathbf{H}^{l}_{i+w}])
\end{equation}
where $\mathrm{Agg}$ is the aggregator function, and $\mathbf{H}^{l}_{j}$ is set to zero if $j$ is less than $0$ or greater than $n-1$.

We propose three different types of aggregators, namely mean aggregator, max aggregator, and attentional aggregator.

\textbf{Mean aggregator}. 
Mean aggregator takes the average of input embeddings. If we treat user behavior sequence as a signal, aggregation layer with mean aggregator performs low pass filtering on the input signal. The high-frequency components in signal, which correspond to interest that changes drastically over time, will be filtered out, leaving relatively stable mid-term or long-term preferences that change slowly. The mean aggregator takes the form:
\begin{equation}
\mathrm{MeanAgg}([\mathbf{H}^{l}_{i-w},\mathbf{H}^{l}_{i-w+1},\cdots,\mathbf{H}^{l}_{i+w}]) = \sum_{j=i-w}^{i+w} \mathbf{H}^{l}_{j}
\end{equation}

\begin{figure}[t!]
	\includegraphics[width=0.48\textwidth]{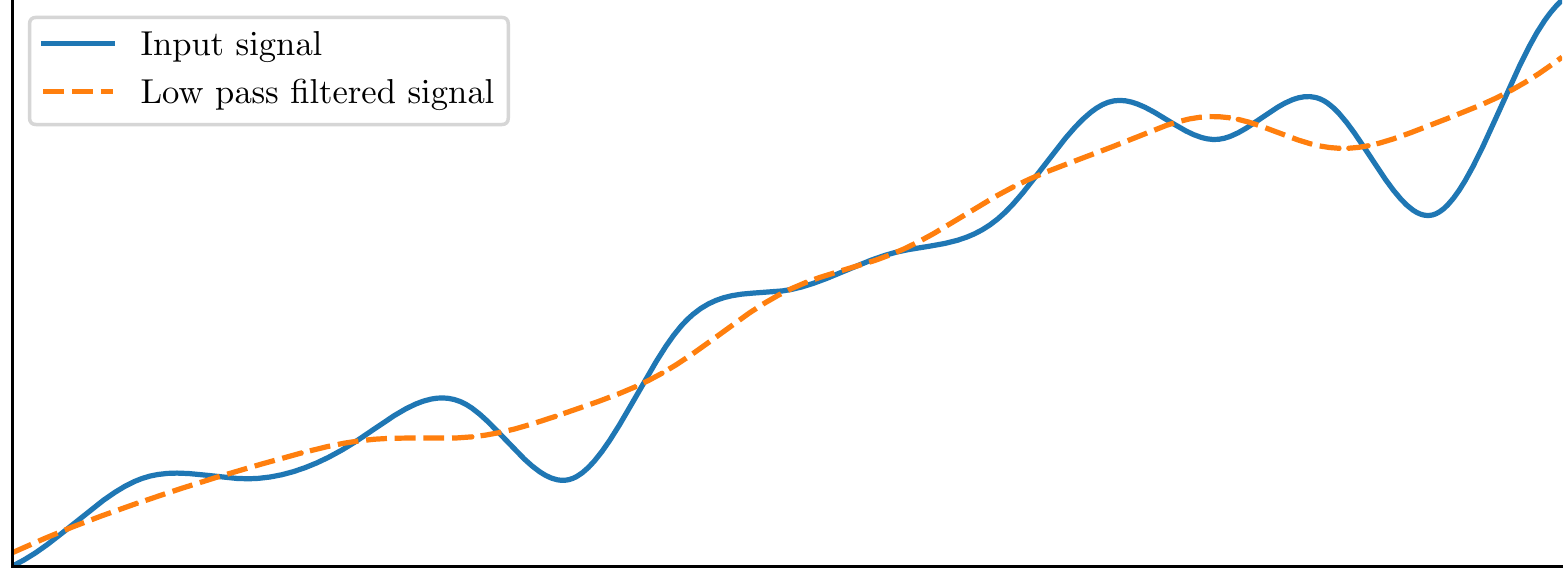}
	\caption{Low pass filtering of a signal. If we regard user behavior sequence as a signal, the high-frequency components correspond to interest that changes drastically, while the low-frequency components correspond to relatively long-term preferences. Average filtering of the user behavior sequence yields interests with longer temporal-ranges.}
\end{figure}

\textbf{Max aggregator}. 
Max aggregator is similar to max-pooling, but instead of selecting the maximum value, max aggregator selects the input embedding with maximum $l2$-norm from input embeddings. The norm expresses the importance of the embedding and thus is used as the indicator to select input embeddings. The max aggregator is as follows:
\begin{equation}
\mathrm{MaxAgg}([\mathbf{H}^{l}_{i-w},\mathbf{H}^{l}_{i-w+1},\cdots,\mathbf{H}^{l}_{i+w}]) = \mathbf{H}^{l}_{\mathrm{argmax}_{j=i-w}^{i+w} \mathrm{norm}(\mathbf{H}^{l}_{j})} 
\end{equation}

\textbf{Attentional aggregator}. 
Attentional aggregator improves over mean aggregator, where each input is weighted by a learned attention score instead of a constant. Attentional aggregator can learn to pay more attention to important parts. The attentional aggregator can be computed as follows:
\begin{equation}
\mathrm{AttnAgg}([\mathbf{H}^{l}_{i-w},\mathbf{H}^{l}_{i-w+1},\cdots,\mathbf{H}^{l}_{i+w}]) = \sum_{j=i-w}^{i+w} \alpha_j \mathbf{H}^{l}_{j} 
\end{equation}
where $\alpha_i$ is the attention parameter related to the position of the embedding which will be learned during training.

\subsection{Attentional Interest Fusion}
The interest aggregation layer produces a group of user interests at different temporal-ranges, attentional fusion structure applies attention mechanism to each interest resolution and then adds them together to form a combined interest representation $\mathbf{h}\in \mathbb{R}^{d}$:
\begin{equation}
\mathbf{h} = \sum_{l} \mathrm{softmax}((\mathbf{H}^l\mathbf{e}_{t+1})^T)\mathbf{H}^l
\end{equation}
where  $\mathbf{e}_{t+1}$ is the embedding of the target item we need to predict. 
We use binary cross entropy loss for training:
\begin{equation}
\mathscr{L} = \sum_{u} -\log(\sigma(\mathbf{h}_u^T\mathbf{e}_u^{+}))-\log(1-\sigma(\mathbf{h}_u^T\mathbf{e}_u^{-}))
\end{equation}
where $\mathbf{e}_u^{+}$ is positive item that is contained in user sequence and $\mathbf{e}_u^{-}$ is randomly sampled negative item that is not in the user sequence.

\section{Experiments}
In this section, we will first describe the datasets, comparing methods, and evaluation metrics, then compare our model against various state-of-the-art recommendation methods on two different datasets and analyze the performance of our method.

\begin{table}
	\caption{Statistics of datasets}
	\begin{tabular}{llllll}
		\hline
		Dataset     & \# Users & \# Items  & \# Actions  \\ \hline
		Electronics      & 11,589    & 20,247          & 347,393 \\
		Movies      & 33,326    & 21,901         & 958,986     \\
		\hline
	\end{tabular}
	\label{tbl:dataset}
\end{table}

\begin{table*}[t!]
	\caption{Performance comparison of different recommendation methods.}
	\small
	\begin{tabular}{llllllllll|lll}
		\hline
		Datasets                 & Metric  & POP    & BPR    & NCF    & DIN    & GRU4Rec & LSTM4Rec & CASER  & SASRec       & MRIF-avg        & MRIF-max & MRIF-attn       \\ \hline
		\multirow{7}{*}{Movie}   & AUC     & 0.7529 & 0.8133 & 0.8233 & 0.8301 & 0.8699  & 0.8700   & 0.8882 & {\ul 0.8960} & 0.8994          & 0.8992   & \textbf{0.9039} \\
		& GAUC    & 0.7532 & 0.8136 & 0.8174 & 0.8295 & 0.8642  & 0.8619   & 0.8847 & {\ul 0.8912} & 0.8948          & 0.8953   & \textbf{0.8980} \\
		& HIT@5   & 0.3200 & 0.4110 & 0.3690 & 0.4730 & 0.4940  & 0.4790   & 0.5380 & {\ul 0.5650} & 0.5660          & 0.5660   & \textbf{0.5720} \\
		& HIT@10  & 0.4560 & 0.5410 & 0.5090 & 0.5910 & 0.6220  & 0.6130   & 0.6790 & {\ul 0.6900} & 0.6940          & 0.6870   & \textbf{0.7060} \\
		& NDCG@5  & 0.2197 & 0.2899 & 0.2576 & 0.3509 & 0.3632  & 0.3542   & 0.3919 & {\ul 0.4248} & 0.4288          & 0.4253   & \textbf{0.4369} \\
		& NDCG@10 & 0.2635 & 0.3324 & 0.3028 & 0.3892 & 0.4050  & 0.3978   & 0.4376 & {\ul 0.4660} & 0.4700          & 0.4646   & \textbf{0.4797} \\
		& MRR     & 0.1130 & 0.1680 & 0.1420 & 0.2160 & 0.2200  & 0.2140   & 0.2350 & {\ul 0.2720} & 0.2760          & 0.2700   & \textbf{0.2870} \\ \hline
		\multirow{7}{*}{Electro} & AUC     & 0.6977 & 0.7568 & 0.7608 & 0.8101 & 0.8491  & 0.8430   & 0.8387 & {\ul 0.8540} & \textbf{0.8565} & 0.8506   & 0.8419          \\
		& GAUC    & 0.6972 & 0.7554 & 0.7606 & 0.8091 & 0.8437  & 0.8394   & 0.8404 & {\ul 0.8493} & \textbf{0.8542} & 0.8506   & 0.8395          \\
		& HIT@5   & 0.2670 & 0.2950 & 0.2890 & 0.3680 & 0.4350  & 0.4070   & 0.4210 & {\ul 0.4430} & 0.4650          & 0.4400   & \textbf{0.4650} \\
		& HIT@10  & 0.3710 & 0.4280 & 0.4220 & 0.5140 & 0.5580  & 0.5500   & 0.5640 & {\ul 0.5780} & 0.5920          & 0.5840   & \textbf{0.5980} \\
		& NDCG@5  & 0.1883 & 0.2049 & 0.1951 & 0.2571 & 0.3000  & 0.2814   & 0.2869 & {\ul 0.3180} & 0.3285          & 0.3138   & \textbf{0.3347} \\
		& NDCG@10 & 0.2217 & 0.2483 & 0.2379 & 0.3045 & 0.3398  & 0.3277   & 0.3330 & {\ul 0.3620} & 0.3700          & 0.3600   & \textbf{0.3774} \\
		& MRR     & 0.1090 & 0.1050 & 0.1020 & 0.1460 & 0.1670  & 0.1500   & 0.1540 & {\ul 0.1820} & 0.1880          & 0.1810   & \textbf{0.1960} \\ \hline
	\end{tabular}
	\label{tbl:performance}
\end{table*}

\subsection{Datasets and Experimental Setup}
We used two datasets, Electronics and Movies, from Amazon dataset in our experiments. Amazon dataset \cite{he2016ups} includes reviews from users on different products. The two subsets used in our experiments have been reduced to extract the 10-cores, such that each of the users in the dataset has at least 10 reviews. The statistics of the two subsets are shown in Table \ref{tbl:dataset}. We use the behavior sequence except the last one of each user for training. For evaluation, the last item of each user is selected as the positive example, and 100 items that are not in user behavior sequence are randomly sampled to serve as negative examples for each user.

\subsection{Compared methods}
The proposed methods are compared against the following baselines: 

\textbf{POP} is item popularity based recommendation method that ignores user-side information.

\textbf{BPR} \cite{rendle2012bpr} uses matrix factorization with pairwise ranking loss.

\textbf{NCF} \cite{he2017neural} augments collaborative filtering with neural networks.

\textbf{DIN} \cite{zhou2018din} uses target item to attend to each historical behavior.

\textbf{GRU4Rec} \cite{hidasi2015gru4rec} applies RNN network with GRU cell to users' historical behaviors.

\textbf{LSTM4Rec} is similar to GRU4Rec except that the RNN cell is LSTM instead of GRU.

\textbf{CASER} \cite{tang2018personalized} leverages CNN networks to capture users' interests.

\textbf{SASRec} \cite{kang2018self} uses self-attention module to model users’ sequential behaviors.

\subsection{Evaluation metrics}
We evaluate model performances in terms of Area under ROC curve (AUC), Group AUC (GAUC), Normalized Discounted Cumulative Gain (NDCG), Hit Ratio (HR), and Mean Reciprocal Rank (MRR). Group AUC (GAUC)  first calculates the AUC within each user, and then computes the sum weighted by the number of samples of each user. HR@k is the fraction of times positive item is ranked among top k, and NDCG@k assigns weight which reduces logarithmically proportional to the position. MRR is the average of the reciprocal ranks.

\subsection{Experimental Results and Analysis}
We show in Table \ref{tbl:performance} the experimental results on two amazon datasets, namely Electronics and Movies. Due to the randomness of algorithms, we perform ten independent runs for each method, and report the average performance. MRIF-avg, MRIF-max, and MRIF-attn are proposed methods with mean aggregator, max aggregator, and attentional aggregator, respectively. We use two aggregation layers with sliding window sizes both set to 3. The best results in compared methods are underlined and the best results among all methods are boldfaced.

The POP method performs worst in terms of all metrics since it only considers the popularity of items, and no user side information is taken into account. BPR and NCF perform better than POP, which is because these two models incorporate user information using collaborative filtering based methods. DIN achieves better results than BPR and NCF on all metrics, since DIN relies on attention mechanism and attends to user's historical behaviors using the target item. GRU4Rec, LSTM4Rec, CASER, and SASRec are all sequential recommendation method which use not only the items that users have interacted with, but also the relative positions of items in sequence. Sequential methods perform better than DIN since the order of items is considered. SASRec outperforms the other three sequential methods with the use of self-attention block. The proposed methods outperform SASRec and achieve best results among all methods. MRIF-attn achieves best results on Movie dataset in terms of all metrics and best results on Electro dataset except under AUC and GAUC metrics, suggesting that the attentional aggregator is the most effective one. MRIF-avg performs slightly worse than MRIF-attn since the weights are constants in mean-aggregator. MRIF-max performs worst among the three proposed methods, which is possibly because that the max-aggregator performs hard aggregation, which only selects one item and thus some auxiliary information is lost.

\section{Conclusions}
In this paper, we propose multi-resolution interest fusion model consisting of interest extraction layer, interest aggregation layer, and attentional fusion structure to address the problem of extracting and combining user preferences at different temporal-ranges. Interest extraction layer relies on transformer blocks to extract instantaneous user interests at each step. Interest aggregation layer focuses on finding a group of user interests at different resolutions. Three different aggregators, which are mean aggregator, max aggregator, and attentional aggregator, are proposed. The interest fusion structure adopts the attention mechanism to integrate multi-resolution interests to make predictions. Experiments on two datasets under seven evaluation metrics demonstrate the superiority of our model.

\bibliographystyle{ACM-Reference-Format}
\bibliography{sample-base}

\end{document}